\documentclass[12pt]{iopart}

\def \BE {\begin{equation}}
\def \EE {\end{equation}}

\def \eq#1 {\eqno{(#1)}}

\def \bl {\mbox{\boldmath{$\ell$}}}
\def \hbl {\mbox{\boldmath{$\hat \ell$}}}
\def \tbl {\mbox{\boldmath{$\tilde \ell$}}}
\def \bn {\mbox{\boldmath{$n$}}}
\def \hbn {\mbox{\boldmath{$\hat n$}}}
\def \tbn {\mbox{\boldmath{${\tilde n}$}}}
\def \bm #1 {\mbox{\boldmath{$m^{(#1)}$}}}
\def \btm #1 {\mbox{\boldmath{${\tilde m}^{(#1)}$}}}

\def \hbm#1 {\mbox{\boldmath{$\hat m^{(#1)}$}}}

\def \BEAH { \begin{eqnarray*}}
\def \EEAH {\end{eqnarray*}}
\def \BEA { \begin{eqnarray}}
\def \EEA {\end{eqnarray}}
\def \BDM {\begin{displaymath}}
\def \EDM {\end{displaymath}}

\begin{document}

\title{A note on the peeling theorem in higher dimensions}
\author{A. Pravdov\' a\dag,  V. Pravda\dag\   and  A. Coley\ddag}
 
\address{\dag\ Mathematical Institute, 
Academy of Sciences, \v Zitn\' a 25, 115 67 Prague 1, Czech Republic}
\address{\ddag\ Dept. Mathematics and Statistics, Dalhousie U., 
Halifax, Nova Scotia B3H 3J5, Canada}

\eads{\mailto{pravdova@math.cas.cz}, \mailto{pravda@math.cas.cz},
\mailto{aac@mathstat.dal.ca}}

\begin{abstract}

We demonstrate the ``peeling property'' of the Weyl tensor in higher dimensions
in the case of even dimensions (and with some additional assumptions), 
thereby providing a first step towards understanding of the general peeling behaviour 
of the Weyl tensor, and the asymptotic structure  
at null infinity, in higher dimensions.

\end{abstract}

\section{Introduction}

The study of higher dimensional manifolds in gravity theory is
currently of great interest. 
A natural question arises which results
concerning four dimensional gravity may be straighforwardly generalized to higher
dimensions. 

Recently the algebraic classification of the Weyl
tensor in higher dimensional Lorentzian manifolds was presented by
characterizing algebraically special Weyl tensors by means of the
existence of aligned null vectors of various orders of alignment \cite{Algclass,CMPP}. This approach
leads to a dimensionally independent classification scheme and reduces to the Petrov
classification in four dimensions (4d).

Now 
it is of interest to determine whether some sort
of peeling theorem is also valid in higher dimensions.
Asymptotic ``peeling properties'' of the Weyl tensor in  physical
$4$-dimensional spacetimes  in 
general relativity  can be preferably  studied within the framework
of conformal null infinity
 \cite{NP,penrose2}, which 
was 
recently 
 introduced also for higher even dimensions
in \cite{holl}
 (see also \cite{wald} for a discussion of odd dimensions).

The purpose of this note is to point out that under certain assumptions the peeling theorem 
is also valid in higher dimensions. Let us emphasize that in 4d the
peeling theorem can be rigorously derived for asymptotically simple spacetimes
using Einstein's equations. 
Here we simply assume that certain asymptotic  properties of the spacetime are satisfied,
which may be true only in particular situations (e.g., even dimensions). 
A more rigorous analysis
of the peeling theorem in higher dimensions would be desirable. However, many necessary related concepts
in higher dimensions are not well understood at present and even the physical importance of the 
concept of conformal infinity in higher dimensions 
is  unclear. 
Perhaps the most promising approach would be to generalize the Bondi method
{ \cite{bondi,sachs2}}
for higher dimensions, but this may require a study of each dimension separately.
This note should  thus be regarded as a first  step in the analysis of the validity of the
peeling theorem in higher dimensions.

\section{Peeling property of the Weyl tensor}

We consider a
$D$-dimensional spacetime \mbox{($M$, $g_{ab}$)}, $D$ even, that is weakly asymptotically simple at null 
infinity \cite{holl}. 
The metric of an unphysical  manifold \mbox{(${\tilde M}$, ${\tilde g}_{ab}$)}  
with boundary  $\Im $, is related to the physical metric by a  conformal transformation
\mbox{${\tilde g}_{ab}=\Omega^2 g_{ab}$,} where
\mbox{$\Omega=0$}  and  \mbox{$   
\Omega_{;a}\not= 0$} is null  at   $\Im $. 
We
note that such a spacetime is vacuum near $\Im $.

We further assume
that components of the unphysical Weyl tensor with respect to the unphysical tetrad 
(see below) are of order
\mbox{${\cal O}(\Omega^{{ q}}) $}  (with \mbox{$q\geq 1 $})
in the neighbourhood of $\Im $. 
In 4d this follows from Einstein's equations (with \mbox{$q=1$}).
At this stage   it is  just a natural assumption from which ``peeling behaviour'' in higher dimensions
follows, and we do not study here to what extent it is  satisfied in general.
One possible method for studying justifiability of this assumption in a given dimension is to use
a generalized Bondi metric \cite{holl} for analyzing asymptotic behaviour of the Weyl tensor.
However, even in the simplest, six dimensional,  case this approach leads to very complicated 
calculations.

Let 
 ${\tilde \gamma}\subset ({\tilde M},\ {\tilde g}_{ab})$ be a null geodesic in the unphysical manifold
that has an affine parameter  ${\tilde r}\sim -\Omega$ near $\Im$ and a tangent vector ${\tbl}$
and    ${ \gamma}\subset ({ M},\ { g}_{ab})$  a corresponding null geodesic in the physical manifold
with an affine parameter   $r\sim 1/ \Omega$ 
near $\Im $ and a tangent vector $\bl$.
In the physical spacetime we will use the frame  $ ( {\bl} ,\ \bn ,\ \bm{i} )$ parallelly propagated along   ${\gamma}$   with respect to ${g}_{ab}$.
$\bn$ and $\bl$ are null vectors
satisfying $\ell^a n_a = 1$, $\bm{i} $ are orthonormal spacelike vectors ($i,j=2\dots D-1$, $a,b,c,d = 0 \dots D-1$).
We choose the corresponding  frame  $(\tbl ,\ \tbn ,\ \btm{i} )$
in the unphysical spacetime
to be related to the physical one by
\BEA
&&{\tilde \ell}_a=\ell_a,\ \ \ \ \ \ \ \ \ 
{\tilde m}^{(i)}_a=\Omega m_a^{(i)},\ \ \ \ \ \ \ \ 
{\tilde n}_a=\Omega^2n_a\ \nonumber\\
\rightarrow &&{\tilde \ell}^a=\Omega^{-2}\ell^a,\ \ \ 
{\tilde m}^{(i)a}=\Omega^{-1} m^{(i)a},\ \ \
{\tilde n}^a=n^a.\label{frameOmega}
\EEA
The physical metric has the form
\BE
g_{ab}= 2 \ell_{(a} n_{b)} + \delta_{ij} m^{(i)}_{a} m^{(j)}_{b} 
\EE
that is preserved by
{null rotations}
\BE
\hbl =  \bl +z_i {\bm{i} } -\frac{1}{2} z^i z_i\, \bn , \ \    
\hbn =  \bn, \ \ 
    \hbm{i} =  \bm{i} - z_i \bn ,
    \label{nullrot}
\EE
 { spins}  and {boosts}
\BDM
\fl
\hbl =  \bl, \ \ \hbn = \bn, \ \hbm{i} =  X^{i}_{\ j} \bm{j} ;\ \ \ \ \ \ \ \ \ \ 
\hbl = \lambda \bl, \ \  \hbn = \lambda^{-1} \bn, \ \ \hbm{i} = \bm{i} .
\EDM
A quantity $q$ is said to have  a  boost weight   $b$   if it transforms under a boost according to
$\hat q = \lambda^b q  $.

Let us now define the operation $\{ \}$ 
$$
w_{\{a} x_b y_c z_{d\}}  \equiv \frac{1}{2} (w_{[a} x_{b]} y_{[c} z_{d]}+ w_{[c} x_{d]} y_{[a} z_{b]})
$$
which allows us to construct    a basis from $ ( {\bl} ,\ \bn ,\ \bm{i} )$ in a vector space of 4-rank tensors with symmetries
$
T_{abcd} = \frac{1}{2} \left( T_{[ab][cd]} + T_{[cd][ab]}\right) .
$
The Weyl tensor  can then be
decomposed in its frame components
with respect to the frame $ ( {\bl}=\bm{1} ,\ \bn=\bm{0} ,\ \bm{i} )$
and these components can be
sorted 
by their boost weight  \cite{Algclass,CMPP}
\BEA
  C_{abcd}&=& 
  \overbrace{
    4 C_{0i0j}\, n^{}_{\{a} m^{(i)}_{\, b}  n^{}_{c}  m^{(j)}_{\, d\: \}}}^2 \nonumber \\
  \nonumber
  &&+\overbrace{
    8C_{010i}\, n^{}_{\{a} \ell^{}_b n^{}_c m^{(i)}_{\, d\: \}} +
    4C_{0ijk}\, n^{}_{\{a} m^{(i)}_{\, b} m^{(j)}_{\, c} m^{(k)}_{\, d\: \}}}^1  
  \nonumber \\
   && \begin{array}{l}
      +4 C_{0101}\, \, n^{}_{\{a} \ell^{}_{ b} n^{}_{ c} \ell^{}_{\, d\: \}} 
\;  + \;  4 C_{01ij}\, \, n^{}_{\{a} \ell^{}_{ b} m^{(i)}_{\, c} m^{(j)}_{\, d\: \}}  \\[2mm]
      +8 C_{0i1j}\, \, n^{}_{\{a} m^{(i)}_{\, b} \ell^{}_{c} m^{(j)}_{\, d\: \}}
   +  C_{ijkl}\, \, m^{(i)}_{\{a} m^{(j)}_{\, b} m^{(k)}_{\, c} m^{(l)}_{\, d\: \}}
    \end{array}
\Biggr\}^0
  \label{eq:rscalars}\\ 
   &&+ \overbrace{
    8 C_{101i}\, \ell^{}_{\{a} n^{}_b \ell^{}_c m^{(i)}_{\, d\: \}} +
    4 C_{1ijk}\, \ell^{}_{\{a} m^{(i)}_{\, b} m^{(j)}_{\, c} m^{(k)}_{\, d\: \}}}^{-1} 
  \nonumber \\
  &&+ \overbrace{
      4 C_{1i1j}\, \ell^{}_{\{a} m^{(i)}_{\, b}  \ell^{}_{c}  m^{(j)}_{\, d\: \}}}^{-2} .\nonumber
\EEA
Components $C_{0i0j}$ have boost weight 2 since they are proportional to $C_{abcd} \ell^a m^{(i)b} \ell^c m^{(j)d}$
and boost weight of other components may be  determined similarly.
Note that the frame components of the Weyl tensor $C_{0i0j},\  \dots, C_{1i1j}$ are subject to a number of constraints
 following from additional symmetries of the Weyl tensor and its tracelessness {\cite{Algclass,CMPP}}.

 Boost order of a tensor ${\mathbf T}$ is defined  as the maximum boost weight of its frame components
and it can be shown that
it depends only on the choice of a null direction  $\bl$ \cite{Algclass,CMPP}.
Boost order of the Weyl tensor in a generic case is 2, but in algebraically special cases, there
exist preferred null directions for which boost order of the Weyl tensor is less. In other words,
in algebraically special spacetimes, one can set all components of boost weight 2, $C_{0i0j}$, to zero 
by an appropriate null rotation (\ref{nullrot}) (we will call this case type I).
Note than in four dimensions this is always possible since the Weyl tensor in 4d always possesses 
principal null directions and thus in 4d type I is generic, while for $D \geq 5$ type I is algebraically special
subclass of the general class G. One can proceed further and say that the Weyl tensor at a given point is of type II, III, and N
if there exists a frame in which boost order of the Weyl tensor is 0, -1, and -2, repectively.

  For spacetimes satisfying the above mentioned assumptions,  
the Weyl tensor's decomposition (\ref{eq:rscalars}) and the relation 
(\ref{frameOmega}) lead to
\newpage
\BEA
  {{\tilde C}_{abc}}\ ^d&=&{\tilde g}^{de} {{\tilde C}_{abce}}={\tilde g}^{de} \biggl[
      4 {\tilde C}_{0i0j}\, {\tilde n}^{}_{\{a} {\tilde m}^{(i)}_{\, b}  {\tilde n}^{}_{c}  {\tilde m}^{(j)}_{\, e\: \}}
 \nonumber \\
  \nonumber
  &&+   8{\tilde C}_{010i}\, {\tilde n}^{}_{\{a} {\tilde \ell}^{}_b {\tilde n}^{}_c {\tilde m}^{(i)}_{\, e\: \}} +
    4{\tilde C}_{0ijk}\, {\tilde n}^{}_{\{a} {\tilde m}^{(i)}_{\, b} {\tilde m}^{(j)}_{\, c} {\tilde m}^{(k)}_{\, e\: \}}
  \nonumber \\
   && 
      +4 {\tilde C}_{0101}\, \, {\tilde n}^{}_{\{a} {\tilde \ell}^{}_{ b} {\tilde n}^{}_{ c} {\tilde \ell}^{}_{\, e\: \}} 
\;  + \;  4 {\tilde C}_{01ij}\, \, {\tilde n}^{}_{\{a} {\tilde \ell}^{}_{ b} {\tilde m}^{(i)}_{\, c} {\tilde m}^{(j)}_{\, e\: \}} 
\nonumber \\&&     
 +8 {\tilde C}_{0i1j}\, \, {\tilde n}^{}_{\{a} {\tilde m}^{(i)}_{\, b} {\tilde \ell}^{}_{c} {\tilde m}^{(j)}_{\, e\: \}}
   +  {\tilde C}_{ijkl}\, \, {\tilde m}^{(i)}_{\{a} {\tilde m}^{(j)}_{\, b} {\tilde m}^{(k)}_{\, c} {\tilde m}^{(l)}_{\, e\: \}}
\nonumber 
\\ 
   &&+
    8 {\tilde C}_{101i}\, {\tilde \ell}^{}_{\{a} {\tilde n}^{}_b {\tilde \ell}^{}_c {\tilde m}^{(i)}_{\, e\: \}} +
    4 {\tilde C}_{1ijk}\, {\tilde \ell}^{}_{\{a} {\tilde m}^{(i)}_{\, b} {\tilde m}^{(j)}_{\, c} {\tilde m}^{(k)}_{\, e\: \}}
  \nonumber \\
  &&+ 
      4 {\tilde C}_{1i1j}\, {\tilde \ell}^{}_{\{a} {\tilde m}^{(i)}_{\, b}  {\tilde \ell}^{}_{c}  {\tilde m}^{(j)}_{\, e\: \}} 
\biggr]\nonumber\\
&=&\Omega^{-2}{g}^{de}  \biggl[
      \Omega^{2+1+2+1} 4 {\tilde C}_{0i0j}\, {n}^{}_{\{a} {m}^{(i)}_{\, b}  {n}^{}_{c}  { m}^{(j)}_{\, e\: \}}
 \nonumber \\
  \nonumber
  &&+   \Omega^{2+2+1}8{\tilde C}_{010i}\, {n}^{}_{\{a} {\ell}^{}_b { n}^{}_c { m}^{(i)}_{\, e\: \}} +
    \Omega^{2+1+1+1}4{\tilde C}_{0ijk}\, {n}^{}_{\{a} { m}^{(i)}_{\, b} { m}^{(j)}_{\, c} { m}^{(k)}_{\, e\: \}}
  \nonumber \\
   && 
      +\Omega^{2+2}4 {\tilde C}_{0101}\, \, { n}^{}_{\{a} { \ell}^{}_{ b} { n}^{}_{ c} { \ell}^{}_{\, e\: \}} 
\;  + \;\Omega^{2+1+1}  4 {\tilde C}_{01ij}\, \, { n}^{}_{\{a} { \ell}^{}_{ b} { m}^{(i)}_{\, c} { m}^{(j)}_{\, e\: \}} 
\nonumber \\&&     
 +\Omega^{2+1+1}8 {\tilde C}_{0i1j}\, \, { n}^{}_{\{a} { m}^{(i)}_{\, b} { \ell}^{}_{c} { m}^{(j)}_{\, e\: \}}
   +  \Omega^{1+1+1+1}{\tilde C}_{ijkl}\, \, { m}^{(i)}_{\{a} { m}^{(j)}_{\, b} { m}^{(k)}_{\, c} { m}^{(l)}_{\, e\: \}}
\nonumber 
\\ 
   &&+\Omega^{2+1}
    8 {\tilde C}_{101i}\, { \ell}^{}_{\{a} { n}^{}_b { \ell}^{}_c { m}^{(i)}_{\, e\: \}} +
    \Omega^{1+1+1}4 {\tilde C}_{1ijk}\, { \ell}^{}_{\{a} {m}^{(i)}_{\, b} { m}^{(j)}_{\, c} { m}^{(k)}_{\, e\: \}}
  \nonumber \\
  &&+ \Omega^{1+1}
      4 {\tilde C}_{1i1j}\, { \ell}^{}_{\{a} {m}^{(i)}_{\, b}  { \ell}^{}_{c}  {m}^{(j)}_{\, e\: \}} 
\biggr]=C_{abc}\ ^d.\nonumber
\EEA      
Since  
all unphysical components of the Weyl tensor
${\tilde C}_{1i1j}$, 
${\tilde C}_{1ijk}$,
${\tilde C}_{101i}$,
${\tilde C}_{ijkl}$,
${\tilde C}_{0i1j}$,
${\tilde C}_{01ij}$,
${\tilde C}_{0101}$,
${\tilde C}_{0ijk}$,
${\tilde C}_{010i}$,
${\tilde C}_{0i0j}$ are assumed to be of order ${\cal O}(\Omega^q)$,
each physical component is of order ${\cal O}(\Omega^{\mbox{{\footnotesize boost weight}}+2+q})$, i.e., 
 we obtain the peelling property 
\BEA
{ C}_{1i1j}&=&{\cal O}(\Omega^q),\nonumber\\
{ C}_{1ijk},\ { C}_{101i}&=&{\cal O}(\Omega^{q+1}),\nonumber \\
{ C}_{ijkl},\
{ C}_{0i1j},\
{ C}_{01ij},\
{ C}_{0101}&=&{\cal O}(\Omega^{q+2}), \  \\
{C}_{0ijk},\ { C}_{010i}&=&{\cal O}(\Omega^{q+3}), \ \nonumber\\
{ C}_{0i0j}&=&{\cal O}(\Omega^{q+4}), \nonumber
\EEA
and thus
\BEA
  C_{abc}\ ^d  =&&  \Omega^q {C^{(N)}}_{abc}\ ^d + \Omega^{q+1} {C^{(III)}}_{abc}\ ^d 
+\Omega^{q+2} {C^{(II)}}_{abc}\ ^d \nonumber\\ &&+\Omega^{q+3} {C^{(I)}}_{abc}\ ^d
\!\!+\Omega^{q+4}{C^{(G)}}_{abc}\ ^d +{\cal O}(\Omega^{q+5} ).
\EEA


We have {thus} shown that from the assumptions outlined above
the ``peeling property'' of the Weyl tensor in the case of even dimensions  follows.
 We hope that this
may provide a first step in proving a peeling theorem in more generality.

\end{document}